# Terahertz Atmospheric Windows for High Angular Resolution Terahertz Astronomy from Dome A

Hiroshi Matsuo[1], Sheng-Cai Shi[2], Scott Paine[3], Qi-Jun Yao[2], Zhen-Hui Lin[2]

[1]National Astronomical Observatory of Japan, Mitaka, Tokyo 181-8588, Japan.
[2]Purple Mountain Observatory, Key Laboratory of Radio Astronomy, Chinese Academy of Sciences, Nanjing 210034, China.
[3]Smithsonian Astrophysical Observatory, Cambridge 02138, Massachusetts, USA.

## Abstract

Atmospheric transmission from Dome A, Antarctica, presents new possibilities in the field of terahertz astronomy, where space telescopes have been the only observational tools until now. Using atmospheric transmission measurements from Dome A with a Fourier transform spectrometer, transmission spectra and long-term stabilities have been analyzed at 1.461 THz, 3.393 THz, 5.786 THz and 7.1 THz, which show that important atmospheric windows for terahertz astronomy open for a reasonable length of time in the winter season. With large aperture terahertz telescopes and interferometers at Dome A, high angular resolution terahertz observations are foreseen of atomic fine-structure lines from ionized gas and a water ice feature from protoplanetary disks.

1. Introduction

Astronomical observations at terahertz frequencies, submillimeter and far-infrared wavelengths between 300 GHz and 10 THz, are key to understanding the origin of our universe and the formation of astronomical objects such as galaxies, stars and planets. With astronomical facilities in space and on the ground, such as the Herschel Space Observatory and the Atacama Large Millimeter/submillimeter Array (ALMA), many terahertz sources are identified and studied in detail. However, in far-infrared or terahertz frequencies above 1 THz, angular resolution is limited due to aperture sizes of the cryogenic space-born telescopes. In the space environment, astronomical interferometry is still a difficult program requiring development of future instrumentation.

Antarctica is another prospective site for terahertz astronomy due to the low precipitable water vapor in the atmosphere. The measurements of atmospheric transmission spectra using Fourier transform spectrometer (FTS) show many atmospheric windows throughout terahertz frequencies (Shi et al. 2016). In this paper we show further analysis of these terahertz windows to confirm their spectral shapes and their long-term stability to examine the feasibility for astronomical observations. Future instrumentation using large terahertz telescopes and interferometry from high altitude sites in

Antarctica is discussed as well.

## 2. Terahertz Astronomy from Antarctica

Astronomical observations in Antarctica have been made from the Amundsen–Scott South Pole station for many years (e.g. Burton 2010). In supra-terahertz frequencies, observation of ionized nitrogen at 1.461 THz was made at the South Pole by Oberst et al. (2011) under observing conditions when atmospheric opacity toward the source was only 3-6%. When observations are made from Dome A, we will be able to observe with higher atmospheric transmission due to much lower precipitable water vapor. For the prospect of further atmospheric windows in terahertz frequencies, we first discuss the astronomical interest of doubly ionized oxygen lines at 3.393 THz and 5.786 THz, and then water ice feature from protoplanetary disks at 7.1 THz.

The doubly ionized oxygen line at 3.393 THz, or [OIII] 88 μm, is attracting interests for observations of very distant galaxies using ALMA. The first successful observation was made toward a galaxy at 13.1 billion light years away by Inoue et al. (2016) revealing the strong [OIII] 88 μm emission line from a low metallicity star-forming galaxy. Recently, Hashimoto et al. (2018) spectroscopically identified the most distant galaxy at redshift 9.11 which is 13.28 billion light years away. There is another line from the doubly ionized oxygen, [OIII] 52 μm, at 5.786 THz which has a different critical density. A ratio of the two lines can be an accurate measure of the ionized gas density, since the far-infrared lines are optically thin throughout the ionized region. These ionized oxygen lines are emitted in far-infrared wavelengths, and it is difficult to observe nearby galaxies due to atmospheric absorption. Only toward high redshift galaxies can the ionized oxygen lines be observed from the ground in submillimeter-wavelengths with ALMA. Nevertheless, observations of the ionized oxygen lines from nearby galaxies with high angular resolution are quite important to understand the physical conditions of massive star-formation in distant galaxies. High-altitude sites in Antarctica provide possible opportunities to observe these ionized oxygen emission lines through atmospheric windows.

Another atmospheric window at 7.1 THz (42 μm in wavelength) is important for studies of exoplanet formation. One of the focuses of exoplanet research is in water-ice features in the middle infrared region as measured in the laboratory by Smith et al. (1994). The Infrared Space Observatory (ISO) has identified this feature in a young star HD14527 (Malfait et al. 1999). The most prominent water ice feature at 43 μm (7 THz) could be observed through the atmospheric window at 7.1 THz, which was a rare occasion to study the water ice feature from the ground. Through the atmospheric window at 7.1 THz, the slope of the ice feature could be observed, and high angular resolution observations will reveal the radial distribution of the feature in different phases; amorphous ice, crystalline ice or liquid water. The scale of these distributions is 1-10 au, corresponding to an angular scale of 10-100 milli-arcsecond (mas) at 100 pc, which could be resolved by astronomical interferometry. The stable atmosphere at Dome A would enable long baseline interferometry at supra-terahertz frequencies, where a baseline of 1 km corresponds to an angular resolution of 10 mas, which is crucial to image the formation sites of earth-like planets around nearby stars.

## 3. Terahertz Atmospheric Windows

Measurements of atmospheric transmission spectra from Dome A were performed using an FTS covering a frequency range 0.75-15 THz in 2010-2011 (Shi et al. 2016). The year-long measurement campaign was aimed at characterizing the atmospheric transmission for future astronomical research and establishing a precise atmospheric model for earth environment studies. An FTS with a

Martin-Puplett type polarizing interferometer was used (Martin and Puplett, 1970). Due to the limited power supply at Dome A, two ambient temperature detectors made of deuterated-triglycine sulfate (DTGS) covered the frequency region 0.75-15 THz. The FTS was installed in an autonomous observatory called PLATO in January 2010 during the 26th Chinese expedition (CHINARE) to Dome A.

The atmospheric transmission spectra have been discussed based on Figure 1 of Shi et al. 2016, which shows whole year and winter time statistics of atmospheric transmission spectra. To discuss the feasibility of astronomical observations through supra-terahertz windows, transmission statistics and long-term stabilities of four terahertz windows of astronomical interest are presented in this paper.

Figure 1 shows one of the best transmission spectra obtained at 12-18h UTC on August 9$^{th}$ 2010, with numerous supra-terahertz windows throughout terahertz frequencies. The supra-terahertz windows at 1.03, 1.3, 1.5 THz, which were identified in Pampa-la-Bola, Atacama (4800 m in altitude) with peak transmittance of 20% (Matsushita et al. 1999), show 50% transmittance from Dome A. A window at 3.4 THz was identified in the FTS measurement from Chajnantor, Atacama (5000 m in altitude) at 10% transmittance (Paine et al. 2000), whereas the Dome A measurement shows transmittance of 25% which is much more suitable for astronomical observations.

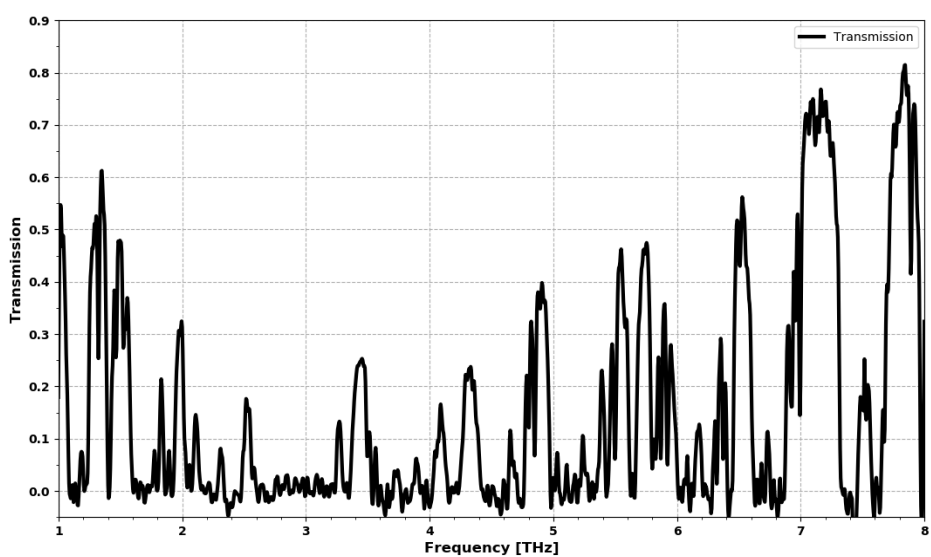

Fig 1. Atmospheric transmission spectrum measured from Dome A at 12-18h UTC on August 9$^{th}$ 2010.

To examine the feasibility of observing the ionized nitrogen line, doubly ionized oxygen lines and ice feature from a protoplanetary disk, expanded views of the four frequencies at 1.461 THz, 3.393 THz, 5.786 THz and 7.1 THz are shown in Figure 2. The stars indicate the frequencies of interest which are actually shifted from the window centers except 7.1 THz (this frequency is selected by the window shape). This results in limitations for observing distant galaxies with redshifted emission lines. For example, a small redshift of up to 0.01 is advantageous for [NII] 1.461 THz observation, but it is difficult to observe galaxies with a redshift around 0.035 due to large H$_2$O absorption at 1.41 THz. Doubly ionized oxygen [OIII] line at 3.393 THz can be observed from nearby galaxies only with a redshift less than 0.01. Another line at 5.786 THz is preferable to observe lower redshift galaxies up to

0.02 and an additional redshift range between 0.03 and 0.05.

Considering the relatively low atmospheric transmittance of supra-terahertz windows, we present a cumulative analysis of the transmittance for the best 10% period in the winter season (April-September), and we expect transmittance better than 21%, 9%, 12% and 56% at 1.461 THz, 3.393 THz, 5.786 THz and 7.1 THz, respectively. Within the four frequencies of interest, the 3.393 THz window shows the lowest transmittance. To check the feasibility of this window, the daily variation in transmittance during the winter season (July-August) is plotted in Figure 3. This shows a transmittance of about 10% continuing for several consecutive days. The transmittance of 10% is not high, but taking into account the stable weather condition and the past [NII] 1.461 THz observations from the South Pole, we think the atmospheric condition at Dome A permit long-term observations through supra-terahertz windows during the winter season.

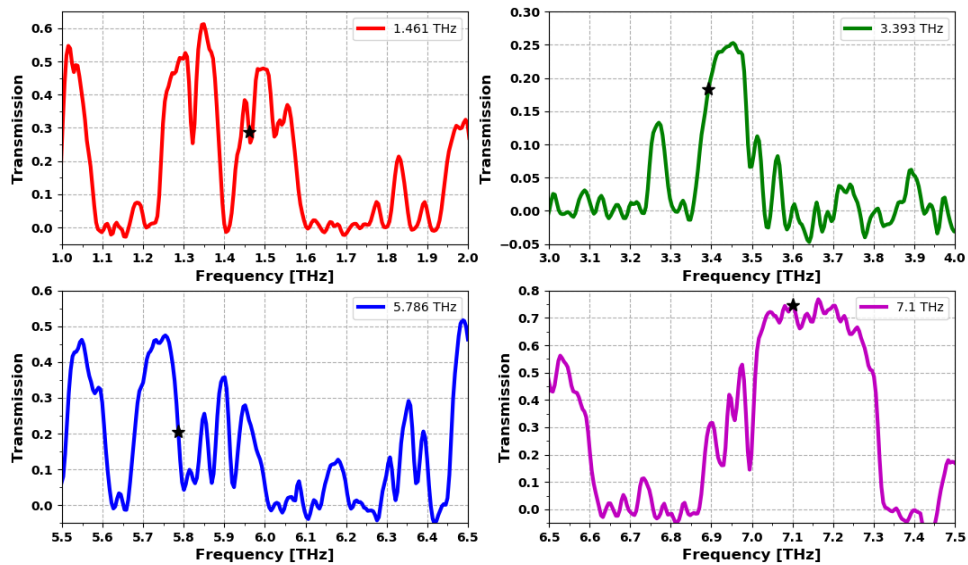

Fig 2. Expanded plot of atmospheric windows at 1.461 THz, 3.393 THz, 5.786 THz and 7.1 THz at 12-18h UTC on August 9th 2010. The stars indicate the frequencies of interest.

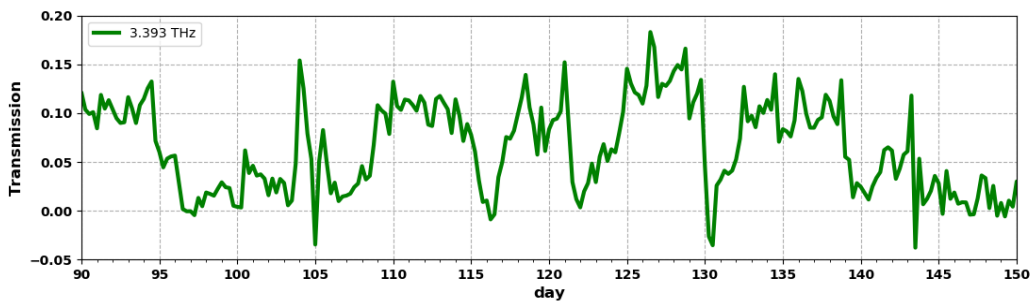

Fig. 3 Long-term stability of atmospheric transmittance at 3.393 THz in the winter season (July-August). For the horizontal axis, 'day' is counted from April 1, 2010. Each data point is an average of 6 hours. The transmittance peak at day 126 is at 12-18h UTC on August 9th 2010.

## 4. Terahertz telescopes and interferometry

Either large aperture telescopes or interferometers can be used for high angular resolution terahertz astronomy. At Dome A, there are two large telescope projects, one optical telescope, KDUST (Zhu et al. 2014) and one terahertz telescope DATE5 (Yang et al. 2013). The ionized nitrogen line at 1.461 THz can be observed with DATE5. At higher frequencies, since efficiency of the DATE5 telescope is not high, only the central part of DATE5 can be used, which will have a similar collecting area to KDUST. To further improve the angular resolution, astronomical interferometry is of interests. Although dedicated interferometry is not easy to install at Dome A, we discuss possibilities of using DATE5 and KDUST in combination with other telescopes to realize terahertz interferometry.

Heterodyne interferometry is a viable option for terahertz interferometry. Superconducting hot electron bolometers with quantum cascade lasers work as supra-terahertz heterodyne receivers, such as demonstrated by Khosropanah et al. (2008) at 3.4 THz. For heterodyne interferometry, phase correction for atmospheric fluctuation will be a key for long baseline observations (ALMA Partnership 2015). Stable atmospheric conditions at Dome A are essential for high angular resolution observation in terahertz frequencies.

Intensity interferometry is another possible option at supra-terahertz frequencies. As demonstrated in the first experiment by Hanbury-Brown and Twiss (1956), intensity interferometry can be used for long baseline interferometers in radio and optical wavelengths. Atmospheric phase fluctuation does not much affect intensity correlations because the correlation length is much longer than a wavelength. Using intensity measurements, it is possible to obtain a delay time using intensity fluctuations observed with two telescopes, which can be used to define complex visibility for aperture synthesis imaging (Matsuo and Ezawa 2016).

For both heterodyne and intensity interferometers, signals can be recorded, and correlation analysis can be made offline as in the case for very long baseline interferometry (VLBI). Image quality of interferometry depends much on the baseline arrangement or, in other words, *uv* coverage. Combining a larger number of telescopes will be advantageous when other telescopes are deployed at Dome A and other high-altitude sites in Antarctica. In general, shorter baseline observations provide astronomical images with larger angular scales and longer baselines give smaller angular scales. The baselines should be arranged according to the astronomical sources of interest. If we increase the baseline length up to 1000 km using a baseline between Dome A and Dome F, 20 micro-arcsecond resolution could be achieved at 3 THz, which provides angular resolution to resolve black holes, stars and exoplanets.

5. Conclusion

Numerous atmospheric windows at Dome A offer new possibilities in future terahertz astronomy, some of which have been discussed in this paper. To show the feasibility of astronomical observations, the statistics of four atmospheric windows, and their long-term stability, were analyzed in winter time. Even for the least transparent window at 3.4 THz, there are many occasions when zenith transmittance of 10 % continues for several days. Astronomy from Antarctica suffers many limitations such as availability of power supply and operational and data transmission difficulties, which are partly similar to satellite environments. Large aperture telescopes and interferometry at Dome A not only opens high angular resolution terahertz astronomy but brings new astronomical findings and technical developments for future space-borne far-infrared and terahertz telescopes and interferometry.

**Acknowledgments**
We acknowledge the assistance of the 26th and 27th CHINARE teams supported by the Polar Research Institute of China and the Chinese Arctic and Antarctic Administration, the University of New South

Wales PLATO team, the Chinese Academy of Sciences Center for Antarctic Astronomy team and the other teams contributing to the operation of the Dome A FTS. H.M. acknowledges Hideko Nomura of Tokyo Institute of Technology for information on the water ice feature of a protoplanetary disk. H.M. was supported partly by a visiting professorship of CAS for senior international scientists.